\documentclass []{aa}
\usepackage{psfig} 
\begin{document}
\thesaurus{03.13.4, 03.13.6, 11.03.1}
\title{Dynamical and content evolution of a sample of clusters from z$\sim$0 
to z$\sim$0.5}

\author{C.~Adami \inst{1,2}, B.P.~Holden \inst{3}, F.J.~Castander \inst{4,3},
A.~Mazure \inst{1}, R.C.~Nichol \inst{5}, M.P.~Ulmer \inst{2}}
                     
\institute{
LAM, Traverse du Siphon, F-13012 Marseille, France  
\and Department of Physics and Astronomy, NU, Dearborn
Observatory, 2131 Sheridan, 60208-2900 Evanston, USA 
\and Department of Astronomy and Astrophysics, University of Chicago,
5640 S. Ellis Avenue, Chicago
\and Observatoire Midi-Pyrenees, 14, Avenue Edouard Belin,
31400 Toulouse, France
\and Department of Physics, Carnegie Mellon University, 5000 Forbes Avenue,
Pittsburgh, PA 15213, USA 
}
\offprints{C.~Adami} 
\date{Received date; accepted date} 

\maketitle 
 
\markboth{Dynamical and content evolution of a sample of clusters from 
z$\sim$0 to z$\sim$0.5}{} 
 
\begin{abstract} 

In this paper, we present an analysis of the dynamics and segregation of 
galaxies in rich clusters from z$\sim$0.32 to z$\sim$0.48 taken from the CFHT 
Optical PDCS (COP) survey and from the CNOC survey (Carlberg et al. 1997). 
Our results from the COP survey are based upon the recent observational work 
of Adami et al. (2000a) and Holden et al. (2000) and use new spectroscopic and 
photometric data on six clusters selected from the Palomar Distant Cluster 
Survey (PDCS; Postman et al. 1996). We have compared the COP and CNOC samples 
to the ESO Nearby Abell Cluster Survey (ENACS: z$\sim$0.07).

Our sample shows that the $z \leq 0.4$ clusters have the same velocity 
dispersion versus magnitude, morphological type and radius 
relationships as nearby ($\sim 0.07$) Abell clusters. The z$\sim$0.48 clusters 
exhibit, however, departures from these relations. Furthermore, there appears 
to be a higher fraction of late--type (or bluer, e.g. Butcher and 
Oemler, 1984) galaxies in the distant clusters compared to the nearby ones.

The classical scenario in which massive galaxies 
virialize before they evolve from late into early type explain our 
observations. In such a scenario, the 
clusters of our sample began to form before a redshift of approximately 0.8
and the late--type galaxy population had a continuous infall 
into the clusters.         

\end{abstract} 
 

\section{Introduction}
                                            
Over the last 25 years, there has been significant effort in
understanding the morphological evolution of galaxies and the dynamical
evolution of the clusters of galaxies. The early work in this area proved 
very fruitful. It
provided strong evidence for galaxy morphological segregation in
clusters (Oemler 1974 and Melnick $\&$ Sargent 1977), {\it i.e.}
early--type galaxies preferentially occupy denser environments and/or
are preferentially found near the cluster centers. In recent years,
there have been many refinements to these initial observations
e.g. Dressler (1980), Kent $\&$ Sargent (1983), Whitmore $\&$ Gilmore
(1991), Biviano et al. (1997), Stein (1997), Adami et al. (1998), Metevier
et al. (2000).

These works have, however, not fully determined whether this
morphological segregation is simply a function of redshift and/or the
cluster environment {\it e.g.} does the galactic content of clusters
evolve with cosmic epoch or as a function of the cluster's virial
mass?  To try to understand these issues, there has been considerable
effort over the last two decades dedicated to studying the dynamical
and morphological evolution of clusters via multi--object
spectroscopy and photometric imaging; for example, Dressler (1980);
Butcher \& Oemler (1984); Gunn et al. (1986); Couch et al. (1991); Bower et
al. (1997); Carlberg et al. (1997); Dressler et al. (1999).

Over the last few years, large surveys of distant clusters of galaxies
-- {\it e.g.}  the Canadian Network of Observational Cosmology (CNOC)
(see Carlberg et al. 1997) -- have begun to provide
interresting answers.  For example, the
red galaxies in the CNOC cluster sample (assumed to be early type galaxies) have a
velocity dispersion nearly $\sqrt 2$ times lower than the blue galaxies 
(assumed to be later type galaxies).

\begin{table*} 
\caption[]{Description of the PDCS clusters sample. The corrected velocity 
dispersions are computed with a decreasing correction of 15$\%$ for the
PDCS clusters.}
\begin{flushleft} 
\begin{tabular}{ccccccc} 
\noalign{\smallskip} 
$los$ & mean z & vel. disp. & corrected vel. disp.& sampling & center (J2000) 
\\ 
\hline 
\noalign{\smallskip} 
PDCS16  & 0.398 & 944$\pm$300 km.s$^{-1}$ & 802 km.s$^{-1}$ & 12 & 
2h28m27.7s 0deg31'58"\\ 
PDCS30/45  & 0.250 & 607$\pm$470 km.s$^{-1}$ & 516 km.s$^{-1}$ & 14 & 
9h54m38.3s 47deg45'40"\\ 
PDCS30/45  & 0.331 & 588$\pm$480 km.s$^{-1}$ & 500 km.s$^{-1}$ & 9 & 
9h54m34.7s 47deg12'10"\\ 
PDCS38  & 0.469 & 1819$\pm$720 km.s$^{-1}$ & 1546 km.s$^{-1}$ & 13 & 
9h51m35.3s 47deg45'40"\\ 
PDCS57  & 0.459 & 1282 km.s$^{-1}$ & 1090 km.s$^{-1}$ & 7 & 
13h23m35.3s 30deg6'10"\\ 
PDCS62  & 0.462 & 1757$\pm$630 km.s$^{-1}$ & 1493 km.s$^{-1}$ & 19 & 
13h23m33.7s 30deg22'48"\\ 
\noalign{\smallskip}  
\hline	    
\normalsize 
\end{tabular} 
\end{flushleft} 
\label{} 
\end{table*}

We have also initiated a new study, the CFH Optical PDCS (COP) 
survey, described in detail in Adami et al. (2000a) and Holden et al. 
(2000).  The COP survey contains approximately 650 galaxy redshift 
measurements (from 6 nights of CFHT time) towards more than 10 PDCS clusters 
at a mean redshift of 0.4. We have also obtained R--band photometry at the 
CFHT for these targets. We have combined our R-band data with the published 
PDCS photometry in V and I. This combined data set is allowing us to study the 
dynamical and morphological status of optically selected distant clusters of 
galaxies. We can then directly compare with the ENACS (ESO Nearby Abell
Cluster survey: e.g. Katgert et al. 1996) and CNOC samples (Carlberg et al. 
1997). Due to the relatively low number of galaxies, the statistics of an 
individual cluster are too low to compare individually to nearby clusters. 
Therefore, we combined the data from all the clusters in our COP study (at 
z$\sim$0.4) whose velocity dispersions were greater than 500 km.s$^{-1}$ into 
one composite cluster. We also applied the same technique to the available 
CNOC data from which we built two composite clusters, one at z$\sim$0.32 and 
one at z$\sim$0.48. Our results and conclusions, then, are based upon the 
comparison of the composite clusters constructed in the same way from the 
ENACS, COP and CNOC surveys. 

In Section 2, we summarize the data and describe the methods used in analyzing 
them. We then discuss evolutionary effects, since our clusters extent over a 
relatively large redshift range and have been all analyzed in the same manner.  
In Section 3, we describe our measurements of galaxy segregation. In Section 4 
we discuss the galaxy evolution in our cluster sample as a function of redshift 
and the implications that our data pose on the formation epoch of these 
clusters. We use q$_0$=0 and $\Lambda$=0. Distances are given in units of 
h=$H_0$/100.

\section{The data}

In this paper we use the COP data described in Adami et al. (2000a) and Holden 
et al. (2000) as well as the ENACS and CNOC data. The 
COP survey is an imaging and spectroscopic follow-up of 10 PDCS (Postman et 
al. 1996) lines--of--sight which contain 15 PDCS cluster candidates at
z$\simeq$0.4 (Adami et al. 2000a and Holden et al. 2000). Here, we
compare the dynamical characteristics of the galaxies belonging to those PDCS clusters to their ENACS (Adami et al 1998) and CNOC counterparts. The low 
redshift data are based on a compilation from the ENACS database (Katgert et 
al. 1996, Mazure et al. 1996) and the literature (Adami et al. 1998). 
All the galaxies in the nearby sample have a redshift and a magnitude that 
were derived in the same manner as the current data. They have also been   
classified into morphological types, but no color data 
comparable to our COP data are available.

We also use 5 CNOC clusters available in the literature (MS0016+16, MS0302+16,
MS1008-12, MS1224+20 and MS0451-03). MS1358+62 and MS1621+26 were excluded
from the sample because of their distorded apparent shape.
The CNOC galaxies have r and g magnitudes and are classified
into spectral types according to their color and spectral distribution (Yee 
et al. 1996). We 
compute an absolute magnitude according to the redshift and the spectral 
type of each galaxy using the k-corrections given in Frei $\&$ Gunn (1994) (see
below for the definition of the CNOC spectral types).

\subsection{Velocity dispersions and definition of the PDCS sample}

We used the most massive ($\geq 10^{14}M_{\bigodot}$) clusters in the PDCS 
sample in our comparison with the ENACS and CNOC surveys in order to match the
same class of clusters. We compute the velocity dispersions of the massive 
systems detected in the COP survey (Adami et al. 2000a) using the bi--weight 
robust estimators of Beers et al. (1990) and applying a 
``3-$\sigma$ clipping''. This is exactly the same methodology used by Adami et 
al. (1998) for the ENACS systems and this ensure us to have robust estimators
despite the relatively low numbers of galaxies used.

Our final velocity dispersion measurements are within 5$\%$ of those presented 
in Holden et al. (2000) for all the clusters in common except PDCS30/45. This 
last group has a complicated structure in redshift space which probably 
produced deviations from Holden et al. in the velocity dispersion 
determination. However, if we 
excluded that cluster from our sample, the results described in the following 
sections do not change by more than 5$\%$. We decided, therefore, to not 
exclude those galaxies when building our composite cluster (see next section).

Due to the small number of redshifts per cluster in our sample ($\sim 10$), we 
were unable to distinguish between inter-loper galaxies -- those which are
passing-through the cluster but are not physically bound to the
cluster -- and real cluster members. Katgert et al. (1996) showed that not 
excluding inter-lopers
results in an increase of the velocity dispersion of the clusters by between 
10$\%$ and 25$\%$. To account for this effect in our sample, we have, therefore,
systematically decreased our measured velocity dispersions by 15$\%$.

For this paper, we selected only the clusters with a redshift
greater than 0.25, a velocity dispersion greater than 500
km.s$^{-1}$ and with more than 7 redshifts (see Table 1). In total, we have 6 
clusters at a mean redshift of 0.4 and 74 cluster galaxies with measured 
redshift and V$_{PDCS}$ and I$_{PDCS}$ magnitudes (56 have also measured R
magnitude). Only one of the clusters in our sample has fewer than 9
redshifts. This is not a problem because the velocity dispersion can be 
reliably estimated using bi-weight estimators with 9 or more redshifts 
(see Lax 1985).

\subsection{The CNOC sample}

We analyzed the CNOC clusters in the same way as the PDCS and 
ENACS clusters. First of all, we defined groups along the lines of sight by 
using the same gap technique as in Adami et al. (1998) or Adami et al. (2000a).
This technique searches for gaps of more than 1000 km.s$^{-1}$ in the  distribution of the galaxies ordered by increasing redshift. The choice of
1000 km.s$^{-1}$ is well adapted to the data we have (see also Katgert et
al. 1996). We checked, however, that values between 600 and 1500 km.s$^{-1}$ 
do not significantly change the results. When more than 5 galaxies were between 
two velocity gaps, we assumed that this was a structure. With this technique 
we detected all the groups defined in Carlberg et al. (1997) with very similar 
mean redshifts (see Table 2) plus several small and probably diffuse groups 
not listed by Carlberg et al.

In order to use a uniform set of selection criteria, we chose only groups with 
a velocity dispersion larger than 500 km.s$^{-1}$. When the detected groups 
were found to be listed in Carlberg et al. (1997), we used the velocity 
dispersion quoted in that paper. For groups not detected by Carlberg et al,
we computed the velocity dispersion by the same method described in the 
previous section. We kept only groups with velocity dispersions larger than 500 
km.s$^{-1}$. In total, only one structure was added to the 5 main structures in 
the CNOC sample (Carlberg et al. 1997). This group, in the MS0016+16 field, is 
at a redshift of 0.328 and has a velocity dispersion of 610 km.s$^{-1}$ 
(denoted MS0016+16$^2$ in the following). Excluding this group did not change 
the results significantly.

Finally, we only used the CNOC galaxies within a radius of 1 h$^{-1}$ Mpc from 
the center in order to match approximately the PDCS and ENACS criteria. For
MS0016+16$^2$, we used the position of the brightest galaxy as the center.

\begin{table} 
\caption[]{Description of the CNOC cluster samples. We have used the 
Carlberg et al. (1997) velocity dispersions, except for MS0016+16$^2$.}
\begin{flushleft} 
\begin{tabular}{ccccccc} 
\noalign{\smallskip} 
$los$ & mean z & vel. disp. & sampling \\ 
\hline 
\noalign{\smallskip} 
MS0016+16  & 0.548 & 1243 km.s$^{-1}$ & 42  \\ 
MS0016+16$^2$  & 0.326 & 610 km.s$^{-1}$ & 8 \\ 
MS0302+16  & 0.425 & 656 km.s$^{-1}$ & 33 \\ 
MS0451-03  & 0.539 & 1354 km.s$^{-1}$ & 43 \\ 
MS1008-12  & 0.307 & 1059 km.s$^{-1}$ & 75 \\ 
MS1224+20  & 0.326 & 798 km.s$^{-1}$ & 24  \\ 
\noalign{\smallskip}  
\hline	    
\normalsize 
\end{tabular} 
\end{flushleft} 
\label{} 
\end{table}

The end result was 6 CNOC structures (see Table 2): three of them at 
z$\sim$0.32 (MS0016+16$^2$, MS1008-12 and MS1224+20); the other three in the 
redshift interval [0.4;0.55] (MS0016+16, MS0302+16 and MS0451-03, mean redshift 
of z$\sim$0.48). We built as described in Section 2.3 two CNOC composite 
clusters with these structures. The z$\sim$0.32 composite cluster has 107 
galaxies and the z$\sim$0.48 composite cluster has 118 galaxies. These numbers
are similar to the number of galaxies in the COP composite cluster.

\subsection{Velocity normalization}

Following Adami et al. (1998), we built "composite clusters" made by combining
the data from all the individual systems. We compute galaxy velocities
relative to the mean velocity and normalized to the velocity
dispersion of each cluster. The normalized velocity of galaxy ''i''
in cluster ''j'', v$_{i_j}$, is defined as

v$_{i_j}$ = (v$_i$-v$_{j_{mean}}$)/$\sigma _j$,

\noindent with v$_{j_{mean}}$ being the mean velocity of 
cluster $j$, $\sigma _j$ being its velocity dispersion and v$_i$ the
original velocity of galaxy ``i''. We merged these normalized 
velocities into the composite clusters. 

The uncertainties could be introduced by systematic errors 
on the $\sigma _j$; however, the use of robust estimators prevents us 
from severe biases in the estimation of the $\sigma _j$. We stress that 
this technique of using composite clusters has been already used successfully,
for example in Adami et al. (1998).

We have also investigated possible bias due to velocity dispersion 
inhomogeneity between the 3 composite clusters. We are not very likely to 
have such selection effects because at z$\sim$0.32, 0.40 and 0.48, the mean 
velocity dispersions of the individual clusters used to make the composite 
clusters are 966 km.s$^{-1}$, 1047 km.s$^{-1}$ and 1119 km.s$^{-1}$ and they 
are overlapping if we consider error bars.
	
\subsection{Estimating the morphological type of the PDCS cluster galaxies}

\begin{table} 
\caption[]{The average V-I, V-R and R-I Colors for the elliptical, the Sbc, 
the Scd and the Im galaxies at 4 different redshifts. These values are taken
from Frei $\&$ Gunn (1994). We have compared our values to these ones to define
an approximate morphological type.}
\begin{flushleft} 
\begin{tabular}{ccccc} 
\noalign{\smallskip} 
color (z) & E & Sbc & Scd & Im \\ 
\hline 
\noalign{\smallskip} 
V$-$I (z=0.25) & 1.69 & 1.21 & 1.00 & 0.78 \\ 
V$-$R (z=0.25) & 0.74 & 0.46 & 0.42 & 0.31 \\
R$-$I (z=0.25) & 0.95 & 0.75 & 0.58 & 0.47 \\\
V$-$I (z=0.30) & 2.00 & 1.28 & 1.08 & 0.81 \\ 
V$-$R (z=0.30) & 0.87 & 0.53 & 0.48 & 0.35 \\
R$-$I (z=0.30) & 1.13 & 0.775 & 0.60 & 0.46 \\
V$-$I (z=0.40) & 2.22 & 1.43 & 1.24 & 0.87 \\ 
V$-$R (z=0.40) & 1.13 & 0.66 & 0.60 & 0.43 \\
R$-$I (z=0.40) & 1.09 & 0.77 & 0.64 & 0.44 \\
V$-$I (z=0.45) & 2.33 & 1.55 & 1.34 & 0.97 \\ 
V$-$R (z=0.45) & 1.12 & 0.73 & 0.64 & 0.49 \\
R$-$I (z=0.45) & 1.21 & 0.82 & 0.70 & 0.47 \\
\noalign{\smallskip}  
\hline	    
\normalsize 
\end{tabular} 
\end{flushleft} 
\label{} 
\end{table}

Before we begin to discuss the morphological content of the PDCS
clusters, we must first establish a common language by which we can
compare our results to past and present work in the field {\it e.g.}
the use of the early and late morphological types.  

Our imaging data did not have the resolution necessary to allow us to
assign galaxies a morphological type. We could, nevertheless, classify
our galaxies in terms of their spectral energy distribution using our
photometry and spectroscopy. Our first classification scheme separates
galaxies into emission line and non-emission line objects.  This
classification relates well to the early/late morphological type
classification (e.g. Biviano et al. 1997). However, in some cases
early type galaxies, like SOs, can exhibit emission lines features in
their spectra. In any case, Biviano et al. (1997) have shown that
less than 4$\%$ of Elliptical or SO cluster galaxies have emission
lines. This implies that our spectroscopically--determined "late type
galaxies" (see Fig. 4) are probably free from significant
contamination by early type objects.

In other cases, some late type galaxy spectra do not have strong emission 
lines, however. Moreover, we used blocking filters to obtain the spectra of 
the COP galaxies (see Adami et al. 2000a) which significantly reduced the 
available spectral range. It is possible, therefore, that we missed emission 
line features due to this reduced spectral coverage. Our "early type galaxies" 
bin may, therefore, be contaminated by late type objects.

The other classification scheme we used was based on the three-color
information (V, R and I) we had for most of the galaxies in the sample
(56 out of 74).  Once the redshift was known, the spectral energy
distribution was constrained and a rough classification between
different spectral types was assigned. We used the Frei $\&$ Gunn
(1994) colors to classify our galaxies into 4 spectral types.  Table 3
gives the V$-$I, V$-$R and R$-$I colors from Frei $\&$ Gunn (1994) for
the 4 different morphological types we used and for the 4 redshifts of our
composite clusters.

We matched the 3 colors of each object with this table.  First, we
transformed the V$_{PDCS}-$I$_{PDCS}$, V$_{PDCS}-$R$_{COP}$ and
R$_{COP}-$I$_{PDCS}$ into the standard Johnson system used in Frei
$\&$ Gunn (1994). As in Postman et al. (1996) and Adami et al
(2000a) we applied the following relations for the color
transformations:

V$-$I = 0.41+0.855(V$_{PDCS}-$I$_{PDCS}$)+ 
0.012(V$_{PDCS}-$I$_{PDCS}$)$^2$,

V$-$R = $-$0.02+(V$_{PDCS}-$R$_{COP}$)$-
$0.056(V$_{PDCS}-$I$_{PDCS}$)+
0.012(V$_{PDCS}-$I$_{PDCS}$)$^2$, and

R$-$I = 0.43+(R$_{COP}-$I$_{PDCS}$)$-$0.089(V$_{PDCS}-$I$_{PDCS}$).

Each color will, in principle, correspond to a single color-type, but
this did not always happen. We therefore developed a method of using the 
colors to uniquely determine the galaxy color-type. First, if all 3 independent 
color-type determinations derived by the colors did yield the same result, we
selected that color-type. This occurred for 24$\%$ of the sample. Second, if 2 
color-types among the 3 were identical, and if the third color-type was close 
to these two (for example, Im is close to Scd), we selected the two identical 
color-types (50$\%$ of the sample). Third, if 2 color-types among the 3 were 
identical, but if the third color-type was not close to these two ones (for 
example, Im is not close to Sbc) we again selected the two identical 
color-types (5$\%$ of the sample). Fourth, if the 3 color-types were different 
and close (for example, E, Sbc and Scd), we selected the mean color-type
(14$\%$ of the sample). Finally, if the 3 color-types were different
but not close (for example, E, Sbc and Im), we selected the color-type
providing the lowest difference with the values of Table 3 (7$\%$ of
the sample).

We have assumed that the $E$ bin of Frei $\&$ 
Gunn (1994) was a mix of elliptical and SO galaxies, the $Sbc$ bin was a mix 
of Sa, Sb and Sc galaxies, the $Scd$ bin was a mix of Sd and Sm galaxies and 
the $Im$ bin contained only irregular galaxies.

This classification technique is similar to the one used
in the CNOC survey (Carlberg et al. 1997), except that we used three colors
instead of two, and we assigned a morphological type (called color-type
hereafter) using 
the Frei $\&$ Gunn (1994) templates. The classification estimates with the 
largest uncertainties are for the last two cases and they constitute only
21$\%$ of the sample. The main uncertainties in the color classification scheme
came from the uncertainties in the magnitudes. Possible dust 
obscuration found, for example, in Cl0039+4713 (z=0.41: Smail et al. 1999) can 
lead to misclassifications, but we ignored this effect. 

We also compare the predicted types of our two classification schemes. 
Overall, emission line galaxies should be mostly late type galaxies and 
absorption line galaxies should be mostly early type galaxies. The
direct comparison shows that, on one hand, 100$\%$ of the emission line 
galaxies of our sample have a color-type later than Sbc (Sbc: 33$\%$, 
Scd:42$\%$, Im: 25$\%$). On the other hand, 14$\%$ of the non-emission 
line galaxies have a color-type later than Scd (e: 50$\%$, Sbc: 36$\%$, 
Scd:0$\%$, Im: 14$\%$). These percentages are well within our expectations 
(see above) and these results give strength to our classification schemes.

However, we must consider this color classification scheme (red
galaxies: early types, blue galaxies: late types) only as an
$indication$ of the true morphological type. The spectral assignment
is clearly not as reliable as the $real$ morphological types obtained,
for example, via HST observations by Dressler et al. (1999) or Lubin
et al. (1998). The morphology and the stellar content of a galaxy can
be affected differently by various astrophysical processes and
consequently the two tentative classifications by colors or spectral
features described in this section must be considered only as an
approximation of the morphological types. However, in order to compare our
distant cluster  with their local counterparts (ENACS)
we needed to establish such a relation between spectral and morphological
classifications.

\subsection{Morphological type of the CNOC cluster galaxies}

In order to relate the CNOC composite clusters with the rest of our
sample (significantly increasing the number of galaxies), we decided
to split the CNOC spectral classification into two bins that could be
related to our previous classification. Hereafter, we refer to these
subsamples as early and late type CNOC galaxies

We assigned the CNOC spectral types 1, 2 (elliptical galaxies) and 3
(E+A galaxies) to the early type class. These are mainly elliptical
galaxies because the CNOC type 3 galaxies represent less than 2.5$\%$
of the sample. The CNOC types 4 and 5 are assumed to be late type
galaxies. In order to compare with the ENACS data, we associated this
late type bin to the bin Sc+Sd+Sm+Irr of Adami et al. (1998).

\section{Evolution of the magnitude and morphological segregation}

As shown in Adami et al. (1998), there exists both a magnitude and a 
morphological segregation of galaxies in nearby clusters that agrees
with simple dynamical models of cluster formation and evolution. 
The bright galaxies have a lower velocity dispersion compared to the fainter 
ones. This variation is expected from energy equipartition in 
clusters. The elliptical galaxies have a lower velocity
dispersion compared to the later morphological types in nearby clusters. 
The ratio of these dispersions between the elliptical and the late type spiral
galaxies is consistent with a model where the late type spirals are
currently falling into the cluster.

\subsection{Magnitude segregation}

We now compare the results of Adami et al. (1998) for nearby clusters
with the results from the 3 distant composite clusters. We
plot in Figures 1, 2 and 3 the variation of the normalized velocity
dispersions versus the absolute magnitude for the 3 distant composite
clusters.  We made 4 bins of 17 distant galaxies for the PDCS
composite cluster at z$\sim$0.4 (Fig. 2), 5 bins of 17 galaxies for
the CNOC composite cluster at z$\sim$0.32 (Fig. 1) and 6 bins of 17
galaxies for the CNOC composite cluster at z$\sim$0.48 (Fig. 3). 

\begin{figure} 
\vbox 
{\psfig{file=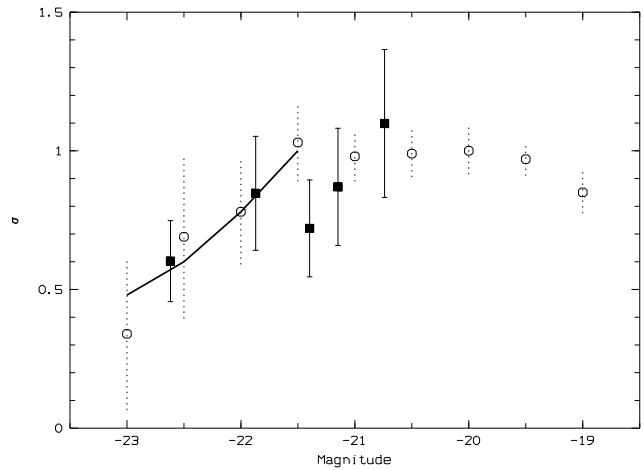,width=9.5cm,angle=270}} 
\caption[]{Magnitude segregation of the 
z$\sim$0.32 and z$\sim$0.07 cluster galaxies. The x-axis is
the R magnitudes in unit of 2.5log$_{10}$(h) + mag., the y-axis is the 
normalized velocity dispersion. The open circles are the data at z$\simeq$0.07
from Adami et al. (1998) with their error bars (dotted lines). The filled 
squares are the data at z$\simeq$0.32 with their error bars (solid lines). 
The thick solid line is the theoretical variation if we 
assume the energy equipartition.} 
\label{} 
\end{figure}

\begin{figure} 
\vbox 
{\psfig{file=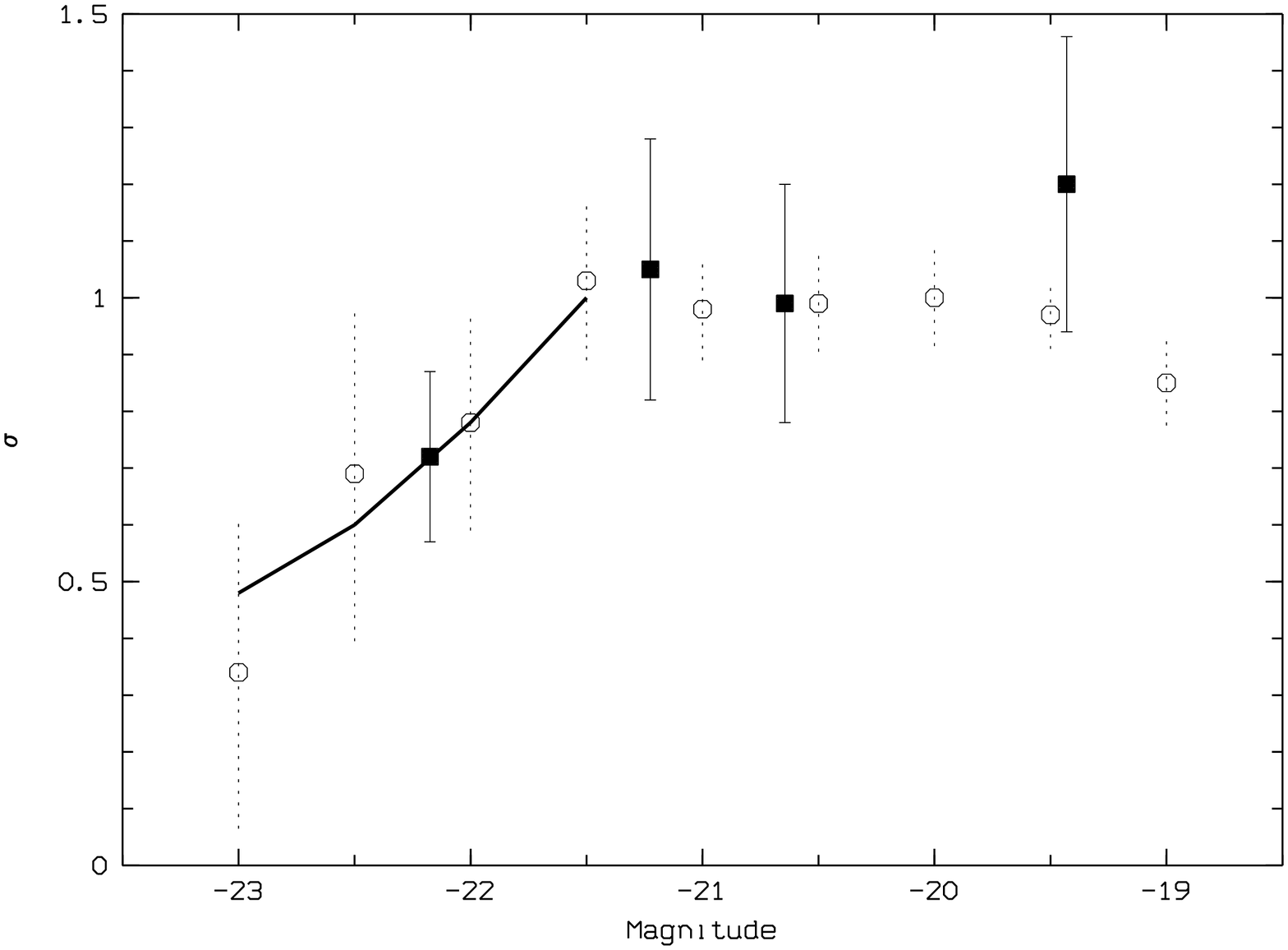,width=9.5cm,angle=270}} 
\caption[]{Magnitude segregation of the 
z$\sim$0.4 and z$\sim$0.07 cluster galaxies. The x-axis is
the R magnitudes in unit of 2.5log$_{10}$(h) + mag., the y-axis is the 
normalized velocity dispersion. The open circles are the data at z$\simeq$0.07
from Adami et al. (1998) with their error bars (dotted lines). The filled 
squares are the data at z$\simeq$0.4 with their error bars (solid lines). 
The thick solid line is the theoretical variation if we 
assume the energy equipartition.} 
\label{} 
\end{figure}

\begin{figure} 
\vbox 
{\psfig{file=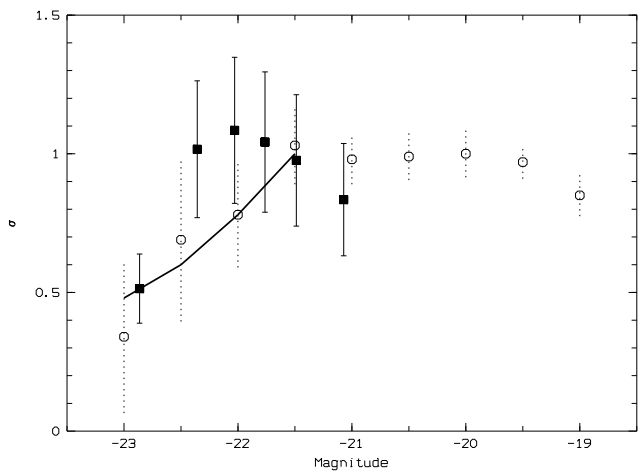,width=9.5cm,angle=270}} 
\caption[]{Magnitude segregation of the 
z$\sim$0.48 and z$\sim$0.07 cluster galaxies. The x-axis is
the R magnitudes in unit of 2.5log$_{10}$(h) + mag., the y-axis is the 
normalized velocity dispersion. The open circles are the data at z$\simeq$0.07
from Adami et al. (1998) with their error bars (dotted lines). The filled 
squares are the data at z$\simeq$0.48 with their error bars (solid lines). 
The thick solid line is the theoretical variation if we 
assume the energy equipartition.} 
\label{} 
\end{figure}

We transformed the V$_{PDCS}$ magnitudes into R magnitudes
(V$_{PDCS}-$R$\simeq$1) to over plot the results for the PDCS
composite cluster on the results for the nearby clusters (which used R
magnitudes). We have not modified the values of the CNOC r magnitudes
because according to Frei $\&$ Gunn (1994), the correction between R
and r magnitudes is less than 0.2 magnitudes. This uncertainty is
negligible compared, for example, to the uncertainty in the
V$_{PDCS}-$R value.

We have excluded galaxies whose velocities were $\geq$3-$\sigma$ away from 
the mean in each bin for the z$\sim$0.4 and z$\sim$0.32 composite clusters.
The figures show qualitatively the same
results of Adami et al. (1998) superposed to each composite
cluster. For the z$\sim$0.32 and z$\sim$0.4 composite clusters, the
agreement is apparently good except for the faintest bin of the
z$\sim$0.4 composite cluster and the R$\sim$-21.5 point of the
z$\sim$0.32 composite cluster (see the discussion section for an
explanation of this peculiar value). For the z$\sim$0.48 composite
cluster, the R$\sim$-22.3 and R$\sim$-22 points seem to be higher than
the corresponding z$\sim$0.07 points as well as the energy equipartition 
prediction (thick solid line).  The energy equipartition law (e.g. Adami et
al. 1998: $\sigma$=10$^{0.2M}$ with M the magnitude) has been
normalized to R = $-$21.5 (+5$\times$log$_{10}$(h)). The agreement
with the low redshift data and the energy equipartition for this last
composite cluster seems to be worse compared to the other two
composite clusters at z$\sim$0.32 and z$\sim$0.4.  However, taking
into account the error bars, the differences are not very significant and
we need to estimate quantitatively the level of agreement.

In order to find a way to quantify more efficiently the possible
differences between the z$\sim$0.07 results and the higher redshift
distributions, we used a two dimensional Kolmogorov-Smirnov test
(e.g. Press et al. 1992) in the normalized velocity dispersion-R
magnitude phase space. We used all the individual values of each galaxy
(without binning) in the magnitude interval R=[-23.;-19.5]. Each
Kolmogorov-Smirnov test was run 1000 times (bootstrap resampling) to
estimate the significance level. At the 10$\%$ level, we found no
significant differences between the normalized velocity dispersion versus R
magnitude for the z$\sim$0.07, z$\sim$0.32 and z$\sim$0.4
composite clusters. However, the distribution of the z=0.48 cluster was found
to be significantly different (even at 1$\%$ level) from the three lower
redshift composite clusters.

The results show, therefore, a significant evolution of the magnitude 
segregation in the most distant composite cluster, but none for the 
z$\sim$0.07, z$\sim$0.32 and z$\sim$0.4 composite clusters.

\subsection{Morphological segregation}

In this section, we also compared the variation of the 
$normalized~velocity~dispersion$ versus the $spectral~types$. Aa a first step, 
we do not consider the spatial distribution of the galaxies (see next Section).

\begin{figure} 
\vbox 
{\psfig{file=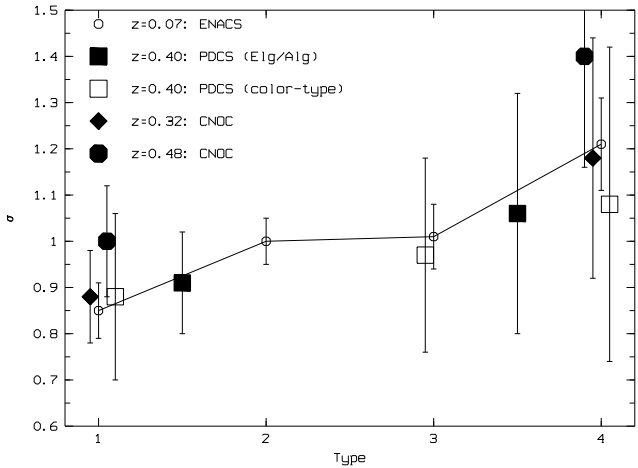,width=9.5cm,angle=270}} 
\caption[]{Morphological segregation. The x-axis is 
the type (1: Ellipticals, 2: SO, 3: Sa+Sb, 4: Sc+Sd+Sm+Irr), the y-axis is the 
normalized velocity dispersion. The linked crossed small circles are the data 
at z$\simeq$0.07 from Adami et al. (1998) with their error bars. The filled
diamonds are the CNOC data at z$\simeq$0.32 with their error bars. The 
large filled circles are the CNOC data at z$\simeq$0.48 with their error bars.
The large squares (filled: emission line/absorption line classification,
opened: color classification) are the PDCS data at z$\simeq$0.4 with their 
error bars. We note that we have applied a small shift in x to the points
at x=1, x=3 and x=4 in order to make the figure clearer.} 
\label{} 
\end{figure}

The types 1, 2, 3 and 4 of Fig. 4 were
respectively assigned to the closest galaxy types: elliptical, SO,
Sa+Sb and Sc+Sd+Sm+Irr.  We stress that the error bars are too
large to draw firm conclusions regarding the morphological
segregations shown in Fig. 4. However, the variation of the mean values
is still suggestive. We now comment on this.

For the z$\sim$0.4 galaxy sample, we used both the emission line (large
filled squares) and color (large empty squares) classifications to
assign morphological types. We used the previously defined early and
late spectral types for the z$\sim$0.32 and z$\sim$0.48 CNOC
samples. The results are shown in Figure 4, together with the
z$\sim$0.07 results (small empty circles) of Adami et al. (1998). To
plot these results on the same figure, we assumed that the
z$\sim$0.4 PDCS galaxies with:

--only absorption lines were an equal mix of elliptical and SO galaxies 

--only emission lines were an equal mix of spiral and irregular galaxies

--a color-type equal to the $E$ type of Frei $\&$ Gunn (1994) corresponded
to type 1 of Adami et al. (1998)

--a color-type equal to the $Sbc$ type of Frei $\&$ Gunn (1994) 
corresponded to type 3 of Adami et al. (1998)

--a color-type equal to the $Scd$ + $Im$ type of Frei $\&$ Gunn (1994) 
corresponded to type 4 of Adami et al. (1998)

Further, we assumed that the CNOC galaxies which are classified as early types 
were Elliptical galaxies and that those classified as late types were type 4 
of Adami et al. (1998)

The results were found to be qualitatively similar for the z$\sim$0.07, 
z$\sim$0.32 and z$\sim$0.4 cluster galaxies. The trend shown is in agreement 
with the theoretical predictions. For example the ratio between the two 
z$\sim$0.32 bins is close to the $\sqrt 2$ value predicted for a relaxed and an 
infalling galaxy population (see e.g. Adami et al. 1998). The ratio is also 
close to the value given in Carlberg et al. (1997).

The results for the z$\sim$0.48 are, however, slightly different
compared to the other redshifts. The ratio of $\sqrt 2$ is still
observed.  There seems to be a systematic shift towards higher values of
the normalized velocity dispersions, however. These distant clusters could be 
dynamically younger compared to the other not so distant clusters.  

\subsection{Spatial distribution segregations}

We now consider the spatial distribution of the galaxies with respect
to their magnitude and spectral type.

\begin{table*} 
\caption[]{Kolmogorov-Smirnov tests regarding the spatial distribution
segregation of the early type and late type galaxies and as a function of the 
R limiting magnitude. A value lower than 0.90 means that the compared samples 
are different at the 10$\%$ level. For a given z bin and a given limiting 
magnitude, the values in the table are the probabilities of the spatial
distribution of the early type galaxies brighter than the limiting magnitude
to be different compared to the spatial distribution of all the late type
galaxies. We give the values only when we have more than 10 early type 
galaxies.}
\begin{flushleft} 
\begin{tabular}{ccccccccccc} 
\noalign{\smallskip} 
z bin/R & -22.75 & -22.5 & -22.25 & -22 & -21.75 & -21.5 & -21.25 & -21 & 
-20.75 & -20.5 \\ 
\hline 
\noalign{\smallskip} 
0.32  & - & 0.69 & 0.65 & 0.83 & 0.98 & 0.98 & 0.95 & 0.98 & 0.98 & 0.99 
\\ 
0.40  & - & - & - & - & 0.25 & 0.65 & 0.95 & 0.95 & 0.95 & 0.91 \\ 
0.48  & 0.79 & 0.94 & 0.99 & 0.99 & 0.99 & 0.99 & 0.99 & 0.99 & 0.99 & 0.99 
\\ 
\noalign{\smallskip}  
\hline	    
\normalsize 
\end{tabular} 
\end{flushleft} 
\label{} 
\end{table*}

We used a two dimensional Kolmogorov-Smirnov test based on the distance
from the center {\it versus} normalized velocity dispersion phase
space (d-$\sigma$ phase space hereafter) to investigate the
segregation as a function of spectral type and magnitude.  In the
z$\sim$0.32, z$\sim$0.4 and z$\sim$0.48 composite clusters, we showed
that the distribution of the early type is significantly different, at
the 10$\%$ level, compared to the late type galaxies. This agrees with
the results of Adami et al. (1998): early type galaxies in nearby
clusters are more concentrated at the cluster center than late type
galaxies.

We also investigated the spatial distribution of the different spectral
types as a function of magnitude using a Kolmogorov-Smirnov test (with
1000 bootstrap resampling for each case). The results are given in
Table 4. For the z$\sim$0.32 cluster, we show that the distribution in
the d-$\sigma$ phase space of early type galaxies brighter than
R$\simeq$-21.75$\pm$0.25 is significantly different at the 10$\%$
level compared to the distribution of late type galaxies. In the same
time, there is no significant difference between late type and early
type galaxies fainter than R$\simeq$-21.75$\pm$0.25. This means that
only the bright early type galaxies are more concentrated at
the center than the late type galaxies. Biviano et al. (2000) found a
similar result in nearby clusters. For the z$\sim$0.4 cluster, we found
the same trend if we used the emission line / absorption line
classification. This time, the segregation appears at magnitudes
brighter than R$\simeq$-21.25$\pm$0.25. For the z$\sim$0.48 cluster,
however, the segregation starts to be significant at
R$\simeq$-22.5$\pm$0.25, brighter than for the
z$\sim$0.32 and z$\sim$0.4 clusters. Once again, the most distant composite
cluster seems to be different form the less distant composite clusters. Only 
the very bright early type galaxies show a more concentrated distribution than 
the late type galaxies.

\subsection{X-ray selected versus optically selected clusters}

The COP clusters are $optically-selected$ while the
CNOC clusters are $X-ray$ selected. We now consider whether these different 
selection methods are responsible for the differences seen between the 
z$\sim$0.48 composite cluster (X-ray selected) and the other composite 
clusters at lower redshift.  For the magnitude and the spatial distribution
segregations, these different selection methods did not seem to play a
significant role because the z$\sim$0.32 (X-ray selected) composite
cluster exhibits the same characteristics as the z$\sim$0.07 and z$\sim$0.4
composite clusters (optically selected) and it was selected in the
same way as the z$\sim$0.48 composite cluster. In the case of the 
morphological segregations, this effect is not statistically significant. We, 
therefore, assume that our results do not depend on whether the clusters were 
X-ray or optically selected.

\section{Discussion}

We stress that our conclusions are strictly valid only for the clusters of 
our sample. However, these clusters are a fair sample of the general 
massive cluster population, and the results presented here can probably be 
generalized to other clusters.

\subsection{How the clusters are evolving?}

Based on our previous analyses we found no evidence for redshift
evolution of the dynamical state in our cluster sample from
z$\sim0.07$ to z$\sim0.4$. We detected, however, significant evolution
between the z$\sim0.4$ and z$\sim0.48$ composite clusters using a
Kolmogorov-Smirnov test.  We also find that the cluster galaxy content
changes from z$\sim$0.07, z$\sim$0.32,
z$\sim$0.4 to z$\sim$0.48. In the nearby sample of cluster galaxies
of Adami et al. (1998: z$\sim$0.07),

--the type ''1'' galaxies (ellipticals) are 24$\%$ of the sample, 

--the type ''2'' (SO) are 49$\%$, 

--the type ''3'' (Sa+Sb) are 20$\%$ and 

--the type ''4'' (Sc+Sd+Sm+Irr) are only 7$\%$. 

In the sample at z$\sim$0.32, 

--the early type galaxies are 78$\%$,

--the late type galaxies are 22$\%$ (type ''4'').

In the sample at z$\sim$0.4, 

--the redder galaxies (assumed to be ''1''+''2'': 
ellipticals + SO) are 44$\%$ of the sample (versus  24$\%$ + 49$\%$ for the
nearby sample), 

--the intermediate color galaxies (assumed to be
type ''3'') are 38$\%$ and 

--the bluer galaxies (type ''4'') are 18$\%$.

And finally, in the sample at z$\sim$0.48,

--the early type galaxies are 70$\%$,

--the late type galaxies are 30$\%$ (type ''4'').

The percentages of late type galaxies are higher in the distant 
cluster samples. The estimate of the uncertainties in these percentages is not 
straightforward but we assumed the Section 2.4 maximum contamination of 21$\%$ 
($\sim$3-$\sigma$) in assigning spectral types when using colors in the 
z$\sim$0.4 sample. To be conservative, we applied this to the relative 
percentages of red, intermediate and blue galaxies of our sample. The 
uncertainties on the numbers given in this paragraph are, therefore, based on 
a 7$\%$ uncertainty.

In other words, the distant cluster galaxies are statistically bluer
(or present later types) than the nearby cluster galaxies. This is the
"classical" Butcher $\&$ Oemler (1984) effect that already has been extensively
reported in the literature. At even higher redshifts, Van Dokkum et
al. (1999) found the same effect in the cluster MS1054-03 at
z$\sim$0.83. They showed clear evidences of further evolution of 
late type galaxies into early types (SO $and$ E).

In order to compare results free from possible bias in our
classification for the z$\sim$0.4 sample, we also compare the
percentage of emission line galaxies in this sample (19$\pm$3$\%$) to
the ENACS sample of nearby clusters (Biviano et al. 1997:
10$\pm$1$\%$), without using any color classification. We see that the
emission line galaxy fraction increases by a factor of $\sim$2. In
principle, our ability to measure redshifts for emission line galaxies
is higher than that for absorption line galaxies, but this effect is
canceled out by the short spectral ranges we used (see Adami et
al. 2000a) which occasionally can lead to missed emission line features. 
The region ($r \sim 1 h^{-1}$ Mpc) covered by the COP
and ENACS surveys are similar (see Adami et al. 2000a and Adami et
al. 1998), which is important as we discuss below.

The evolution in the galaxy content of the clusters in our sample between
z$\sim$0 and z$\sim$0.48 seems to be real. It does not
depend on the sample used to test it, whether it is selected according
to their spectral features or to their colors.
These results are consistent with the recent studies of 
Dressler et al. (1999) of z $\sim 0.4 $
clusters. They find evolution of the morphological types of 
cluster galaxies.

The metric aperture used, however, plays a key role in
determining the blue fraction: For example, the implied evolution of
the blue fraction with redshift was not confirmed by Lubin et al. (1998)
and Postman et al. (1998). They showed that the massive cluster
1604+4304 at z=0.84 has a morphological content similar to that of 
nearby clusters.  But, these studies used images of the inner core of
the cluster ($\sim 2' \times 2'$) which is significantly smaller than
the area sampled in Adami et al. (1998), our comparison study. If we
assume that the percentage of late type galaxies increases from the
inner part to the outer parts of the clusters (e.g. Adami et
al. 1998), then this is an explanation of the discrepancy between our
results and the results of Lubin et al. (1998) and Postman et
al. (1998).  In order to check this point, we have computed the
percentage of red, intermediate and blue galaxies in our z$\sim$0.4
sample but in the same physical area as in Lubin et al. (1998): the
core of the clusters. We found in this region, 65$\%$ red galaxies,
18$\%$ intermediate galaxies and 17$\%$ very blue galaxies. These
percentages are in agreement with the work of Lubin et al. (1998).
The choice of metric aperture is therefore important, and it could be
an explanation for why the longer look--back time studies of Lubin and
Postman did not yield evidence for evolution. It may be that
at longer look back times (z$\sim$0.8) other effects could lead to a
relatively low blue galaxy fraction (see section 5).

The entire evolutionary process is complicated, however, and it probably also
relates to the dynamical state of the clusters as well as its redshift. For
example, Wang et al. (1997) showed a correlation between cluster ellipticity
and blue galaxy fraction, and Metevier et al. (2000) showed that there exist
nearby clusters that have bi-modal galaxy or X-ray distributions which have 
relatively high blue fractions and are yet at low ($\sim 0.1$) redshifts.

\subsection{Morphological evolution versus dynamical evolution}

When we compared the dynamical and the cluster content evolution of our 
composite clusters between z$\sim$0.4 and z$\sim$0.48, the situation is
consistent: we both found "younger" clusters and a
"younger" population at higher redshift. However, the
composite clusters from z$\sim$0.07 to z$\sim$0.4 give the following 
puzzle: How did the cluster contents evolve in morphological type, while the 
clusters have not apparently evolved in terms of their dynamical state?  We 
now examine this point.  

The answer is, it is possible to have the bright galaxies virialize before 
changing
morphological type. This would explain why our composite nearby and distant
clusters have the same distribution of galaxies versus magnitude, but they do 
not have the same ratio of late to early type galaxies. We can see this by
assuming that galaxies virialize via dynamical friction, which takes 0.1 Gyrs 
for the brightest galaxies for example in our $z\sim 0.4$ sample and 1 to 1.5 
Gyrs for the faintest galaxies (Sarazin, 1986).  Yet, if galaxy-galaxy 
interactions cause the galaxies to change morphological type, the galaxies
take  about 1 Gyr to change (crossing time), and clearly the more massive 
(brighter) galaxies can reach  a ``relaxed state'' before changing 
morphological type. Without additional assumptions, however, this scenario runs 
into a difficulty; namely, the time between $z\sim 0.07$ and $z\sim 0.4$ is 
about 4 Gyrs, and the nearby clusters should have virialized such that the 
dynamical state of the nearby and distant clusters should not be the same.

A way to deal with this is to assume that clusters are (not surprisingly) 
continually evolving
(e.g. Ulmer et al 1992 and references therein). This scenario is clearly not
new and has been investigated in details by authors such as Oemler (1974), 
Melnick $\&$ Sargent (1977), Dressler (1980) or Whitmore $\&$ Gilmore (1991). 
However, here, we have a unique case to check this scenario with homogeneously 
analyzed clusters from z$\sim$0 to z$\sim$0.5. Clusters are evolving being 
fed continuously by the in-fall of field galaxies and groups (or other 
clusters). The majority of the infalling field galaxies are late-type. The most 
recent 
episode of in-fall could have been such that the relative low mass late
type galaxies in the outer regions have been able to evolve into early 
type galaxies ($\sim$1 Gyr) while yet not virializing ($\sim$1.5 Gyr). This 
leads to the nearby clusters 
having higher velocity dispersions in their outer portions while, at the same 
time, having relatively low fractions of late-type galaxies. 

This hypothesis is consistent with finding a correlation with dynamical age and
blue fraction (Wang et al. 1997). It is not completely clear how this 
hypothesis relates to finding nearby clusters which have bi-modal distributions 
and appear to have relatively large blue fraction (Metevier et al. 2000). In a 
high matter density Universe, however, the cluster formation occurs at late 
epochs (close to the present time) and a high density Universe could then leads 
to the production of dynamically young structures until relatively recent
epochs.

If we assume the general picture of cluster evolution and formation, we can 
extrapolate this scenario back in time to place a 
lower limit on when clusters formed and also suggest what clusters should look 
like in terms of morphology and dynamical state if we are able to extend 
studies to redshifts up to $\sim 0.8$. This is discussed in the next 
subsection.

\subsection{Cluster formation epoch?}
             
An interesting question to ask is: when did all these clusters form? This epoch 
depends strongly of the cosmological model (e.g. Oukbir $\&$ Blanchard 1992 and 
1997). In a high $\Omega _m$ Universe, the clusters form later than in a low 
$\Omega _m$ Universe, but they evolve faster. To adress this question, we can 
use the composite clusters of our sample from z$\sim$0.07 to z$\sim$0.4. These
are already, at least partially, relaxed. We can then assume that the 
"formation" process has occurred at an epoch earlier than z$\sim$0.4. The 
clusters of the z$\sim$0.48 sample have younger dynamical characteristics and 
we can assume that they are close to the $end$ of their "first formation" 
process, 
which probably occured around z$\sim$0.5. Now we have to estimate the time 
needed to form such a cluster to estimate the epoch of the $beginning$ of the 
formation process.

The surface covered by the clusters of our samples is a circle of about 
$\sim$1h$^{-1}$ Mpc radius. The minimal time to form such a cluster is then
less than 1 Gyear in a violent relaxation process (1 crossing
time, see Sarazin 1986).  Adding 1 Gyear to z$\sim$0.5 gives the epoch of the
beginning of the cluster formation, i.e. in a q$_0$=0 Universe $\sim$7.2 
Gyears (not significantly different from when we use low values of q$_0$ but 
different from 0). We then conclude that the cluster formation process for the 
clusters of our sample began at z$\geq$0.8.

Consistent with this hypothesis are the recent detections of clusters of 
galaxies at z$\sim$0.8 (in addition to the two quoted in section 4.1: 
MS1054-03 and 1604+4304). Nichol et al. (1999) and Ebeling et al. 
(1999) report the discovery of an X-ray luminous cluster (RXJ0152) at 
z=0.83, Gioia et al. (1999) has also discovered a young X-ray luminous cluster 
at z=0.81 and Donahue et al. (1998) reported one at z=0.82. There is also 
the detection of an X-ray cluster (Rosati et al. 1999) at z=1.26 with a low 
X-ray luminosity. Nearly all these clusters at z$\geq$0.8 have a disturbed X-ray
shape (elongated and clumpy) and their nature (partially relaxed clusters or 
proto-clusters) is not clear yet. The formation epoch of these massive 
clusters of 
galaxies, though, was probably only slightly greater than z=0.8, which is 
consistent with our estimates of the minimum z of cluster formation. Selection 
effects may also have played a role in finding only distorded clusters (cf. 
Adami et al. 2000b), however, and the cluster formation may occur at even
higher redshifts (cf. Rosati et al. 1999) for (at least) the low mass clusters. 

\section{Summary and conclusions}

We have produced different composite clusters based on combining the results of 
6 rich optically selected clusters with redshifts near 0.4, and 6 X-ray 
selected clusters with redshifts near 0.3 and 0.5.  We compared these to a
similar composite cluster at redshift 0.07. We have shown that in terms of 
brightness, type and radius versus velocity dispersion, the $z\sim 0.07$, 
$\sim 0.32$ and $\sim 0.4$ composite clusters are very similar (at least within 
the statistical limitations of the data) while the $\sim 0.48$ composite
cluster is significantly different. We conclude from this work that the 
clusters of the sample have at least partially virialized by a redshift of 
0.5 and that 
massive clusters ($\sim$10$^{15}$ solar masses) have formed at the very latest 
at z = 0.8.  Our derived lower limit to the time of cluster formation epoch is 
consistent with work on finding massive clusters at z = 0.8 or higher.  

We have confirmed that the more distant clusters of our sample contain more 
late-type galaxies
that nearby clusters, as found many times before.  At the same time, however,
we have shown that the dynamical state of these clusters does not seem to
evolve for the redshifts lower than 0.4. In order to explain this, 
to the extent that there are some bright blue galaxies in the core of the 
clusters, we hypothesize that these massive blue galaxies have virialized 
before changing morphological classification.  This is not a large effect, 
however, as the cores of the clusters are mainly red (cf. Section 4.1).  We 
must, therefore, conclude (at the level of statistics we have here) that it is 
primarily the outer regions of the clusters that contribute the relatively 
large blue fractions of distant clusters. Although the nearby clusters should 
have virialized completely, the outer portions of these clusters are not 
virialized and yet are dominated by red galaxies.  Continuous in-fall of 
relatively numerous, low mass galaxies such that the 
galaxies have evolved into early type galaxies is just consistent with the 
time scales involved (1Gy for relaxation for low mass or faint galaxies and 
1.5 Gy for virialization).

It is not clear what our proposed scenario predicts at greater look back times.
One possibility is that early type galaxies formed first and formed the 
initial clusters (this hypothesis is, however, not supported by van Dokkum et 
al. 1999). Then the later type galaxies, having formed later {\em in the field} 
are just beginning to fall into the cluster and even the bright, massive ones 
have not yet fallen in. In this scenario, the cluster core will have a low blue 
fraction/low fraction of late type galaxies even at the bright end. This is 
approximately consistent with the finding of Wirth et al. (1994) on CL 0016+16 
(albeit an X-ray luminous cluster) or of Butcher $\&$ Oemler (1984) or as
suggested by Lubin et al (1998) and Postman et al (1998) (see, however, 
section 4.1 regarding the Lubin et al. 1998 and Postman et al. 1998 results) 

The detailed scenario we suggested to explain the results of our analysis and
comparisons with a nearby optically selected sample should be considered only
that, a suggestion.  More elaborate theoretical work, that is beyond the scope 
of this paper, is necessary to develop the full range of possibilities. The
statistics of the results we found here also clearly need to be improved by 
analyzing the data from more clusters with larger telescopes. What our 
discussion has showed is the power of the analysis we performed here, and how 
more analysis of $z\geq 0.5$ systems coupled with an extension to even higher 
redshift systems will do much to clarify picture of how galaxy populations 
evolve within clusters.

\begin{acknowledgements}

{AC thanks the staff of the Dearborn Observatory for their hospitality 
during his postdoctoral fellowship. The authors thanks the CFHT TAC for
support for the COP survey. }

\end{acknowledgements}

\vfill 

\begin{thebibliography}{} 

\bibitem{}  Adami, C., Ulmer, M.P., Romer, A.K., et al. 2000b, ApJ, submitted
\bibitem{}  Adami, C., Holden, B.P., Castander, F., et al. 2000a, AJ, submitted
\bibitem{}  Adami, C., Biviano, A., $\&$ Mazure, A. 1998, A$\&$A, 331, 439
\bibitem{}  Beers, T.C., Flynn, K., $\&$ Gebhardt, K. 1990, AJ, 100, 32
\bibitem{}  Biviano, A., et al. 2000, A$\&$A, in preparation
\bibitem{}  Biviano, A., Katgert, P., Mazure, A., et al. 1997, A$\&$A, 321, 84
\bibitem{}  Bower, R.G., Castander, F.J., Ellis, R.S., Couch, W.J., $\&$ 
Boehringer, H. 1997, MNRAS, 291, 353
\bibitem{}  Butcher, H., $\&$ Oemler, A. 1984, ApJ, 285, 426
\bibitem{}  Carlberg, R.G., Yee, H.K.C., Ellingson, E., et al. 1997, ApJ, 476, 
7
\bibitem{}  Couch, W.J., Ellis, R.S., MacLaren, I., Malin, D.F. 1991, MNRAS,
249, 606
\bibitem{}  van Dokkum, P.G., Franx, M., Fabricant, D., Kelson, D.D., $\&$ 
Illingworth, G.D. 1999, ApJ, 520, L95
\bibitem{}  Donahue, M., Voit, G.M., Gioia, I.M., et al. 1998, ApJ, 502, 550
\bibitem{}  Dressler, A., Smail, I., et al. 1999, ApJS, 110, 213  
\bibitem{}  Dressler, A. 1980, ApJ, 236, 351
\bibitem{}  Ebeling, H., Jones, L.R., Perlman, E., et al. 1999, ApJ, submitted
astroph: 9905321
\bibitem{}  Frei, Z., $\&$ Gunn, J.E. 1994, AJ, 108, 4
\bibitem{}  Gioia, I.M., Henry, J.P., Mullis, C.R., et al. 1999, AJ, 117, 2608
\bibitem{}  Gunn, J.E., Hoessel, J.G., Oke, J.B. 1986, ApJ, 306, 30
\bibitem{}  Holden, B.P., et al. 2000, AJ, submitted
\bibitem{}  Katgert, P., Mazure, A., Perea, J., et al. 1996, A$\&$A, 310, 8
\bibitem{}  Kent, S.M., $\&$ Sargent, W.L.W. 1983, AJ, 88, 697
\bibitem{}  Lax, D. 1985, J. Am. Stat. Assoc., 80, 736
\bibitem{}  Lubin, L.M., Postman, M., Oke, J.B., et al. 1998, AJ, 116, 584
\bibitem{}  Mazure, A., Katgert, P., den Hartog, R., et al. 1996, A$\&$A, 310, 
31
\bibitem{}  Melnick, J., $\&$ Sargent, W.L.W. 1977, ApJ, 215, 401
\bibitem{}  Metevier, A., Romer, A.K., Ulmer, M.P. 2000, AJ, in press
\bibitem{}  Nichol, R.C., Romer, A.K., Holden, B.P., et al. 1999, ApJ, 521, 21
\bibitem{}  Oemler, A.Jr. 1974, ApJ, 194, 1
\bibitem{}  Oukbir, J., $\&$ Blanchard, A. 1997, A$\&$A, 317, 1
\bibitem{}  Oukbir, J., $\&$ Blanchard, A. 1992, A$\&$A, 262, L21
\bibitem{}  Postman, M., Lubin, L.M., Oke, J.B. 1998, AJ, 116, 560
\bibitem{}  Postman, M., Lubin, L.M., Gunn, J., et al. 1996, AJ, 11, 615
\bibitem{}  Press, W.H., Teukolsky, S.A., Vetterling, W.T., Flannery, B.P. 1992,
Numerical Recipes: Second Edition
\bibitem{}  Rosati, P., Stanford, S.A., Eisenhardt, P.R., et al. 1999, AJ, 118, 
76
\bibitem{}  Sarazin, C.L. 1986, Rev. Mod. Phys., 58, 1
\bibitem{}  Smail, I., Morrison, G., Gray, M.E., et al. 1999, ApJ, 525, 609
\bibitem{}  Stein, P. 1997, A$\&$A, 317, 670
\bibitem{}  Ulmer, M.P., Wirth, G.D., Kowalski, M.P. 1992, ApJ, 397, 430
\bibitem{}  Wang, Q.D., Ulmer, M.P., Lavery, R.J. 1997, MNRAS, 288, 702
\bibitem{}  Whitmore, B.C., $\&$ Gilmore, D.M. 1991, ApJ, 367, 94
\bibitem{}  Wirth, G.D., Koo, D.C., $\&$ Kron, R.G., 1994, ApJ, 435, L105
\bibitem{}  Yee, H.K.C., Ellingson, E., Abraham, R., et al. 1996, ApJS, 102, 289
\end{thebibliography}
\end{document}